# Utilization of 3D segmentation for measurement of pediatric brain tumor outcomes after treatment: review of available free tools, step-by-step instructions, and applications to clinical practice

Marina Kazarian[1], Sandra Abi Fadel MD[2], Amit Mahajan MD[2], Mariam Aboian MD/PhD[3].
[1]Horace Mann School, Bronx, NY, [2]Yale New Haven Hospital, New Haven, CT, [3]Yale University - MEDDRA Radiology, New Haven, CT

**Yale Department of Radiology**
**Contact Information:**
Mariam Aboian, MD/PhD
Section of Neuroradiology and Nuclear Medicine
Department of Radiology, Yale University School of Medicine
mariam.aboian@yale.edu

---

**Abstract**

Volumetric measurements are known to provide more information when it comes to segmenting tumors, in comparison to one- and two-dimensional measurements, and thus can lead to better informed therapy. In this work, we review the free and easily accessible computer platforms available for conducting these 3D measurements, such as Horos and 3D Slicer and compare the segmentations to commercial Visage software. We compare the time for 3D segmentation of tumors and demonstrate how to use a novel plugin that we developed in 3D slicer for the efficient and accurate segmentation of the cystic component of a tumor.

---

**Introduction**

Measurement of response assessment in solid tumors relies on several different methods such as one dimensional measurements (RECIST), two dimensional measurements (RANO), and volumetric measurements[1-4]. Prior research suggested that one- and two-dimensional measurements correlate to volumetric measurements, but recent literature in pediatric brain tumors suggests that there is an added value to volumetric measurements[5]. The major drawback in clinical implementation of volumetric measurement of pediatric brain tumors is the time it takes to segment the tumor and lack of widespread availability of 3D measurement tools among different PACS systems. We present how to obtain volumetric measurements using 3D measurement tools available to radiologists for free and the time it takes to segment with each platform in order to determine the most efficient one. We also describe a novel plug-in for 3D slicer for measurement of cystic component within tumor. We compare this plug-in to a similar tool in Horos that also returns accurate volumetric measurements of the cystic tumor component. We show the segmentation of pilocytic astrocytomas in pediatric patients using these tools. The segmentations are saved as NIfTI – extension .nii - files.

**Methods**

We measured the tumor volumes of 5 pediatric patients with pilocytic astrocytomas using volumetric measurement tools such as Visage 3D segmentation (A), 3D Slicer (B), and Horos (C). Step by step methods are shown below on how to perform the segmentations are available in attachments. For 3D slicer, we used both standard segmentation techniques (2) and a new module (3) that automates the segmentation of ROIs based on varying intensity levels surrounding a point on the tumor.

Figure 1A: Visage

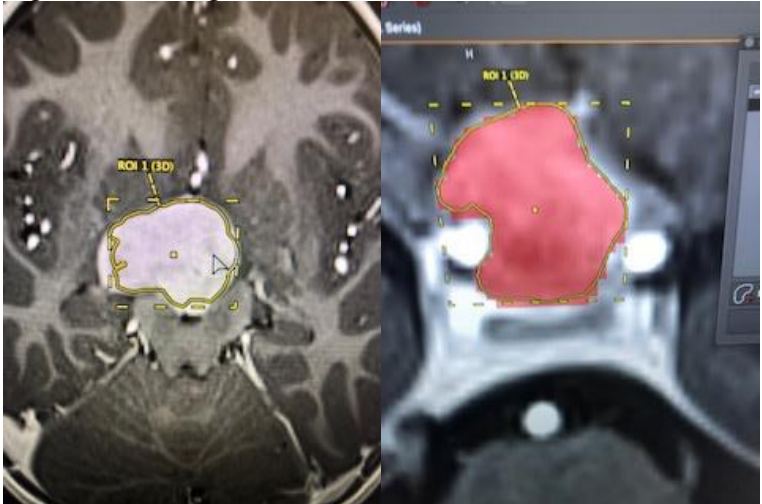

Within the computer platform Visage, select New 3D under structures and segment the tumor using contouring palette. Select ROI slices and create selection for the slice contours. Edit the structure, create new selection and save the segmentation. The segmentation can now be exported into a .nii file.

Figure 1B: 3D Slicer

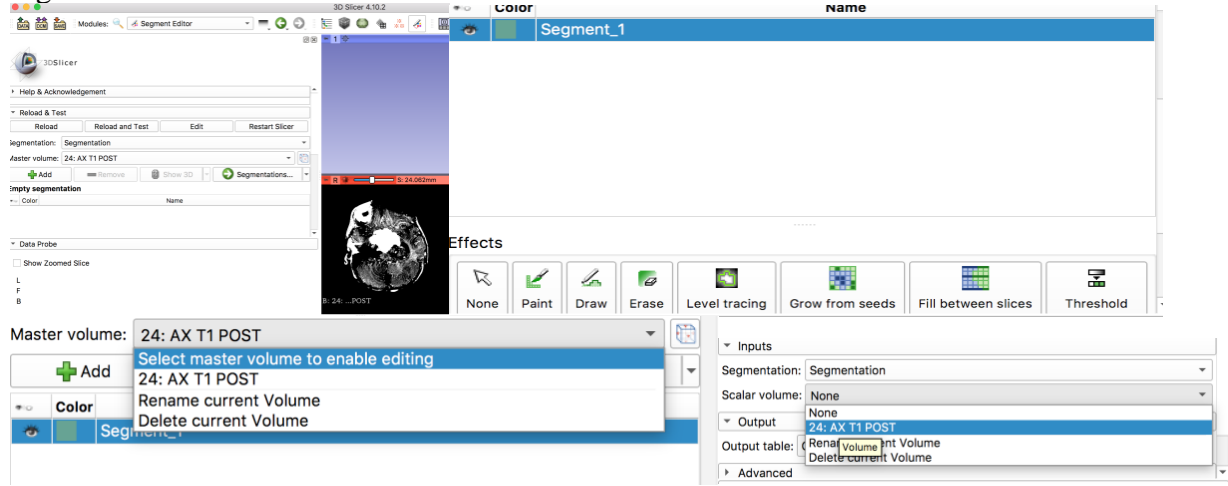

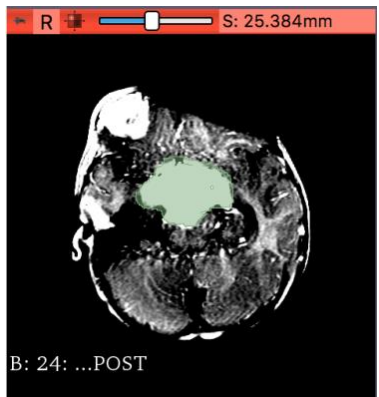

Within the 3DSlicer platform, load DICOM data and select the scan you want to use. Select the segment editor tool from the overhead module panel and add a segmentation. Then, in the master volume dropdown menu, select your scan in order to be able to draw new segmentations and obtain their volumes. Choose the effect to use (eg paint) and draw in ROIs, click right and left arrow keys to change slices. Import/export your segmentation and choose to export it to a new labelmap, which will later be saved as a NIfTI file. In the modules dropdown, select segment statistics under quantification, specify input image and apply to get a table with the volume. Then, go to save in the overhead panel and select only the files that have extension .nrrd, which should be the scan and the exported segmentation, change the extension to .nii, and save them. You can also save the table that was generated with the volumetric measurements.

Figure 1C: 3D Slicer/Intra-tumor cyst volume

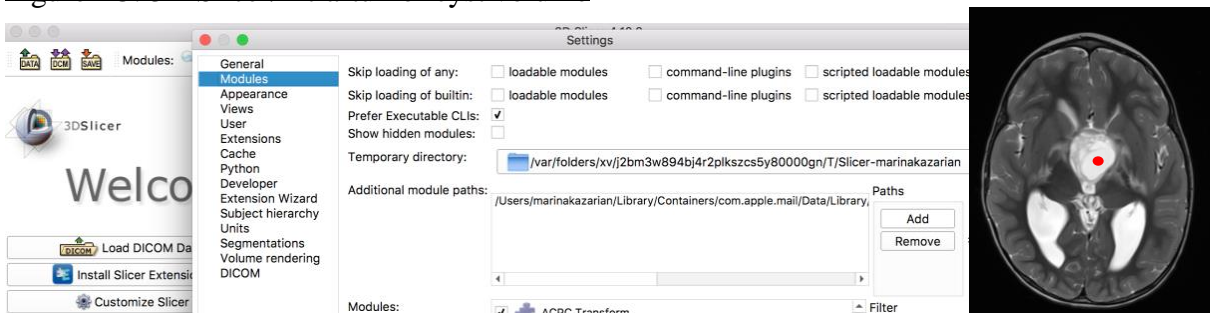

With our new module in 3D Slicer, you can obtain intra-tumor cyst volume. To get the folder with the code for this module, please contact us. In 3D Slicer, click Customize Slicer and add a new module under the modules tab. Select the folder with the code for cystic volume estimation to add as the new module. After loading DICOM data as in [1B], use the fiducial tool with the blue arrow and red dot from the module panel at the top of the screen to place points on the cyst within the tumor on T2 weighted imaging. Search for the new module name with the magnifying glass, press calculate volume and open the python interactor (the rightmost tool) to display the volume of the cystic tumor component. Not only will the module return the volume, but it also creates a segmentation of the area around the fiducial within the borders of the cystic tumor component. The module calculates this area based on intensity, or brightness, of the area. So, if the segmentation created highlights too much or too little of the desired portion, adjust the intensity slider and re-calculate. Before re-calculating, be sure to delete existing segmentations under the clean-up tab within the module in order to ensure that the new segmentation created is not overlapping another preexisting one. To save the segmentation as in [1B], export the segmentation under the segment editor tool to a new labelmap. Since a new segmentation was generated, if you go to segment statistics under quantification as in [1B] a table with the volume of the segmentation can be created and then saved.

Figure 1D: Horos

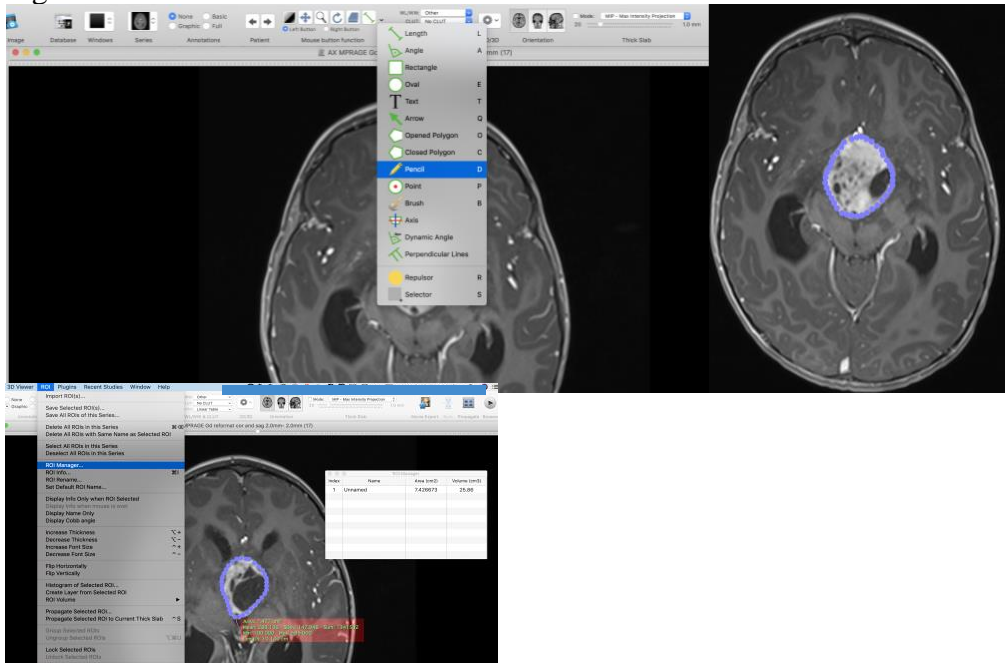

In Horos, select the study and bring the image of post-gadolinium T1 weighted imaging on full screen. Select drawing tool under the tool palette and draw the margins of the tumor on all slices. Under ROI tab, select all ROI and then select ROI manager. The manager will demonstrate the volume of the segmented tumor.

Figure 1E: Horos/Intra-tumor cyst volume

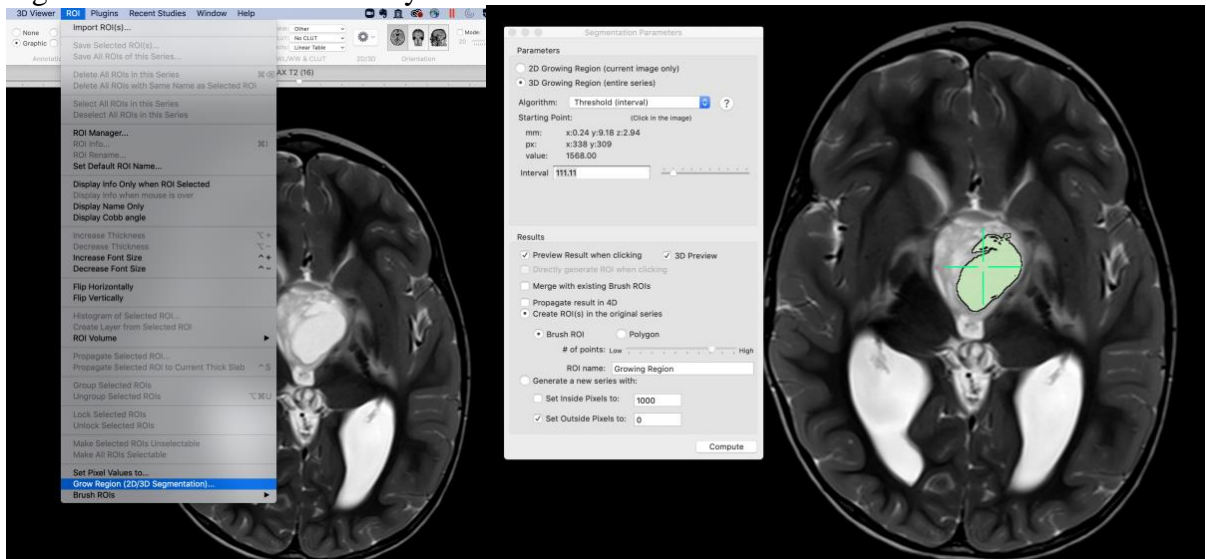

Tumor cyst volume can be measured on Horos on T2 weighted images by selecting 2D/3D Segmentation growing region from the ROI dropdown and putting one dot in the middle of the largest cyst. Select the 3D growing region option and compute the volume of the cystic portion.

## Results
Table 1:

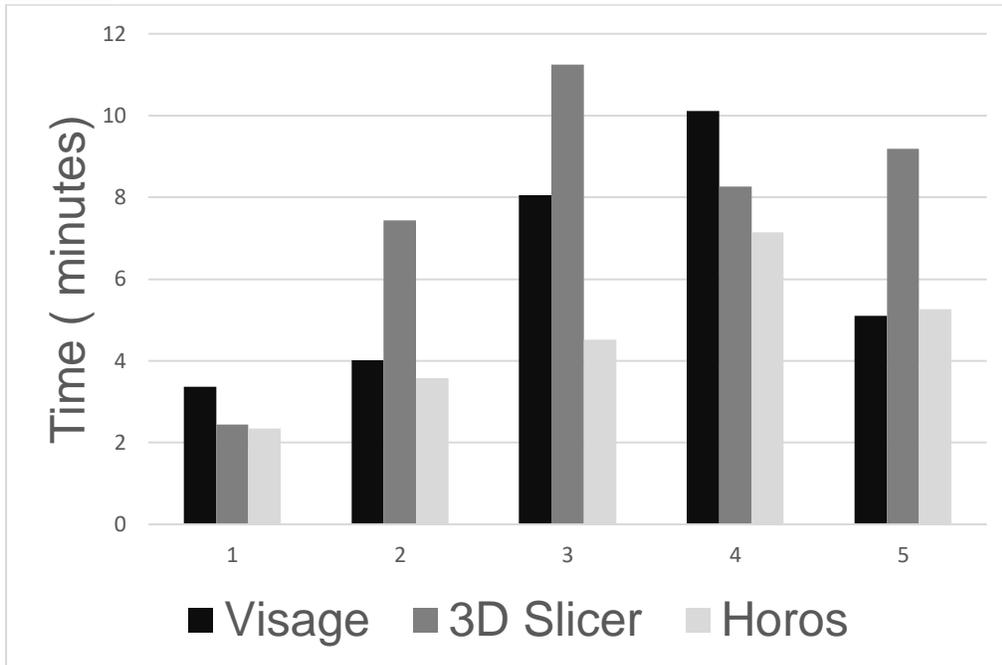

Volumetric measurement of tumors is a time intensive process, although incorporation of 3D measurement tools into PACS systems has decreased the time it takes for segmentation. We measured the time it took for experienced neuroradiologist to measure tumor volume using different tools including Visage, 3D Slicer, and Horos on five different pediatric pilomyxoid astrocytomas.

Figure 2:

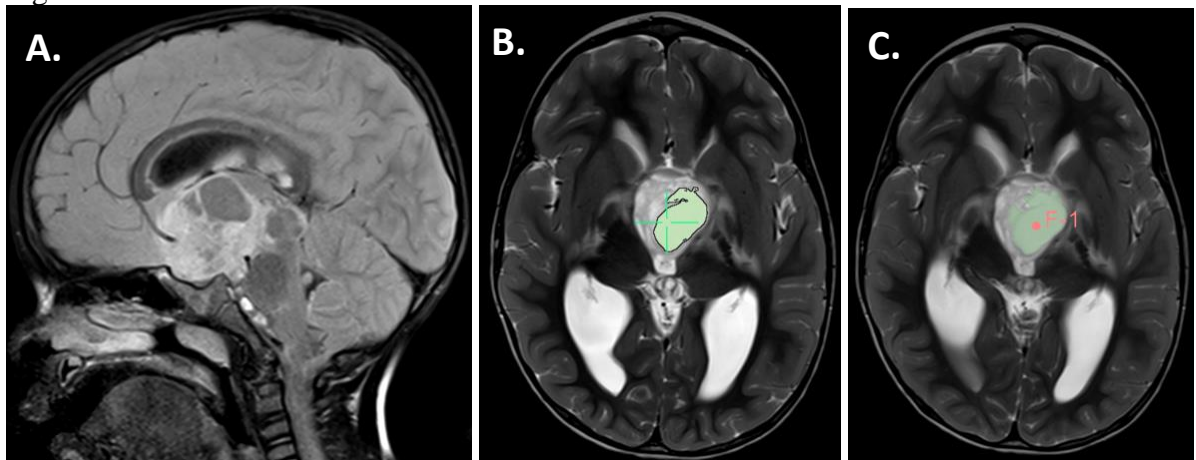

|  | Horos | 3D Slicer Plug-in |
|---|---|---|
| **Total Volume (mm$_3$)** | 28923 | 28923 |
| **Cystic Volume (mm$_3$)** | 3356 | 3710 |
| **Solid Volume (mm$_3$)** | 25567 | 25213 |
| **% Cystic** | 11.60% | 12.83% |

Volumetric measurement of solid and cystic portions of the suprasellar/hypophyseal region pilomyxoid astrocytoma using Horos and 3D Slicer plug in. (A) Sagittal FLAIR demonstrates a large heterogeneous mass with cystic and solid components. Evaluation of tumor response to treatment is usually performed by measuring the AP x trans x CC dimensions, but with 3D segmentation tools, volume of total mass, solid component, and cystic component is possible. (B) Segmentation of the cystic component using Horos. (C) Segmentation of the cystic component using 3D Slicer plugin.

**Conclusions**
The availability of intuitive, quick, and reliable volumetric assessments can lead to qualitatively better therapeutic outcomes. We demonstrated two easily available and free PACS systems that provide 3D measurements of brain tumors and measurement of % cystic component of the tumor. Our research shows that creating new freely available modules to further automate tasks improves efficiency compared to other semi-automatic processes.

**References**

1. Eisenhauer EA, Therasse P, Bogaerts J, et al. New response evaluation criteria in solid tumours: revised RECIST guideline (version 1.1). *Eur J Cancer* 2009; **45**(2): 228-47.
2. An YY, Kim SH, Kang BJ, Lee AW, Song BJ. MRI volume measurements compared with the RECIST 1.1 for evaluating the response to neoadjuvant chemotherapy for mass-type lesions. *Breast Cancer* 2014; **21**(3): 316-24.
3. Berntsen EM, Stensjoen AL, Langlo MS, et al. Volumetric segmentation of glioblastoma progression compared to bidimensional products and clinical radiological reports. *Acta Neurochir (Wien)* 2020; **162**(2): 379-87.
4. Aras M, Erdil TY, Dane F, et al. Comparison of WHO, RECIST 1.1, EORTC, and PERCIST criteria in the evaluation of treatment response in malignant solid tumors. *Nucl Med Commun* 2016; **37**(1): 9-15.
5. D'Arco F, O'Hare P, Dashti F, et al. Volumetric assessment of tumor size changes in pediatric low-grade gliomas: feasibility and comparison with linear measurements. *Neuroradiology* 2018; **60**(4): 427-36.